\documentclass{amsart}

\usepackage{fullpage, amsthm, epsfig,amssymb}

\newcommand{\be}{\begin{equation}}
\newcommand{\ee}{\end{equation}}
\newcommand{\bea}{\begin{eqnarray}}
\newcommand{\eea}{\end{eqnarray}}
\newcommand{\beas}{\begin{eqnarray*}}
\newcommand{\eeas}{\end{eqnarray*}}

\def\bkw{\;\raisebox{-8mm}{\epsfysize=24mm\epsfbox{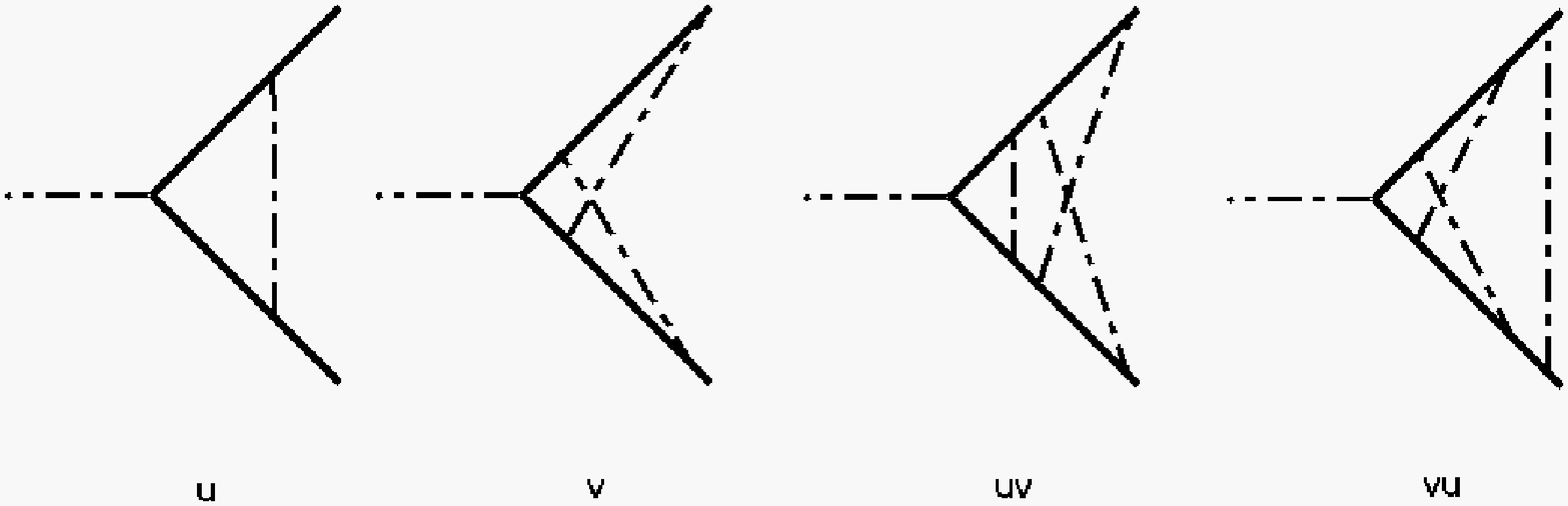}}\;}

\newtheorem{thm}{Theorem}

\def\One{\mathbb{I}}

\title{The next-to-ladder approximation for linear Dyson-Schwinger equations}
\author{Isabella Bierenbaum$^{1}$, Dirk Kreimer$^{2}$ and Stefan Weinzierl}
\address{isabella.bierenbaum@desy.de, DESY, Zeuthen; kreimer@ihes.fr, IHES (http://www.ihes.fr) and Boston U.\ (http://math.bu.edu);
stefanw@thep.physik.uni-mainz.de, ThEP, Institut f\"ur Physik, Universit\"at Mainz}
\thanks{${}^1$supported by DFG Sonderforschungsbereich Transregio 9, Computergest\"utzte Theoretische Physik.}
\thanks{${}^2$supported by CNRS and in parts by grant NSF-DMS/0603781.}
\begin{document}
\setlength{\baselineskip}{0.515cm}
\sloppy
\thispagestyle{empty}
\begin{flushright}
DESY 06--233 \hfill\\
SFB/CPP--06--58\\[2em]
\end{flushright}

\maketitle
\section{Introduction}
Ladder approximations have been one of the most basic attempts to
simplify and truncate Dyson--Schwinger equations in field theory in
a still meaningful way \cite{george}. From a mathematical viewpoint
they simplify the combinatorics of the forest formula considerably,
and are solvable by a scaling Ansatz for sufficiently simple
kinematics.

Here, we discuss such a scenario, but iterate one- and two-loop
skeletons jointly, combining some analytic progress with a thorough
discussion of the underlying algebraic properties.
\subsection{Purpose of this paper}
The main purpose is to sum an infinite series of graphs based on the
iteration of two underlying skeleton graphs. We progress in a manner
such that our methods can be generalized to any countable number of
skeletons. We restrict to linear Dyson Schwinger equations, a case
relevant for theories at a fixpoint of the renormalization group. We
proceed using one-dimensional Mellin transforms, a privilege of
linearity of which we make full use. See \cite{BKDSE,dktor,KY,KY2} for the general approach.
\section{The Dyson--Schwinger Equation}
\subsection{The integral equation}
The equation which we consider is in massless Yukawa theory in four-dimensional Minkowski space $\mathbb M$, for
pedagogical purposes. We define a renormalized Green function describing the coupling of a scalar particle to
a fermion line by
\bea G_R(a,\ln (-q^2/\mu^2)) & = &
 1-a\int_{\mathbb M} \frac{d^4k}{i \pi^2}
 \left\{ \frac{1}{k\!\!/}G_R(a,\ln (-k^2/\mu^2))\frac{1}{k\!\!/}\frac{1}{(k-q)^2}\right\}_- \nonumber\\
 & & + a^2 \int_{\mathbb{M}} \frac{d^4k}{i \pi^2} \int_{\mathbb{M}} \frac{d^4l}{i \pi^2}
 \left\{\frac{l\!\!/(l\!\!/+k\!\!/)G_R(a,\ln (-(k+l)^2/\mu^2))(l\!\!/+k\!\!/)(k\!\!/+q\!\!/)}{[(k+l)^2]^{2}l^2(k+q)^2k^2(l-q)^2}\right\}_-
 , \label{DSEint}
\eea
 where $\{\}_-$ indicates subtraction at $\mu^2=-q^2$,
so that $G_R(a,0)=1$:
\bea
 \left\{ G_R(a,\ln (-k^2/\mu^2))\right\}_{-}
        = G_R(a,\ln (-k^2/\mu^2))-G_R(a,\ln (-k^2/-q^2)).
\eea
The kinematics are such that the fermion has momentum $q$ and the external scalar particle carries zero momentum.
The equation can graphically be represented as
\\
\\
{\epsfxsize=160mm\epsfbox{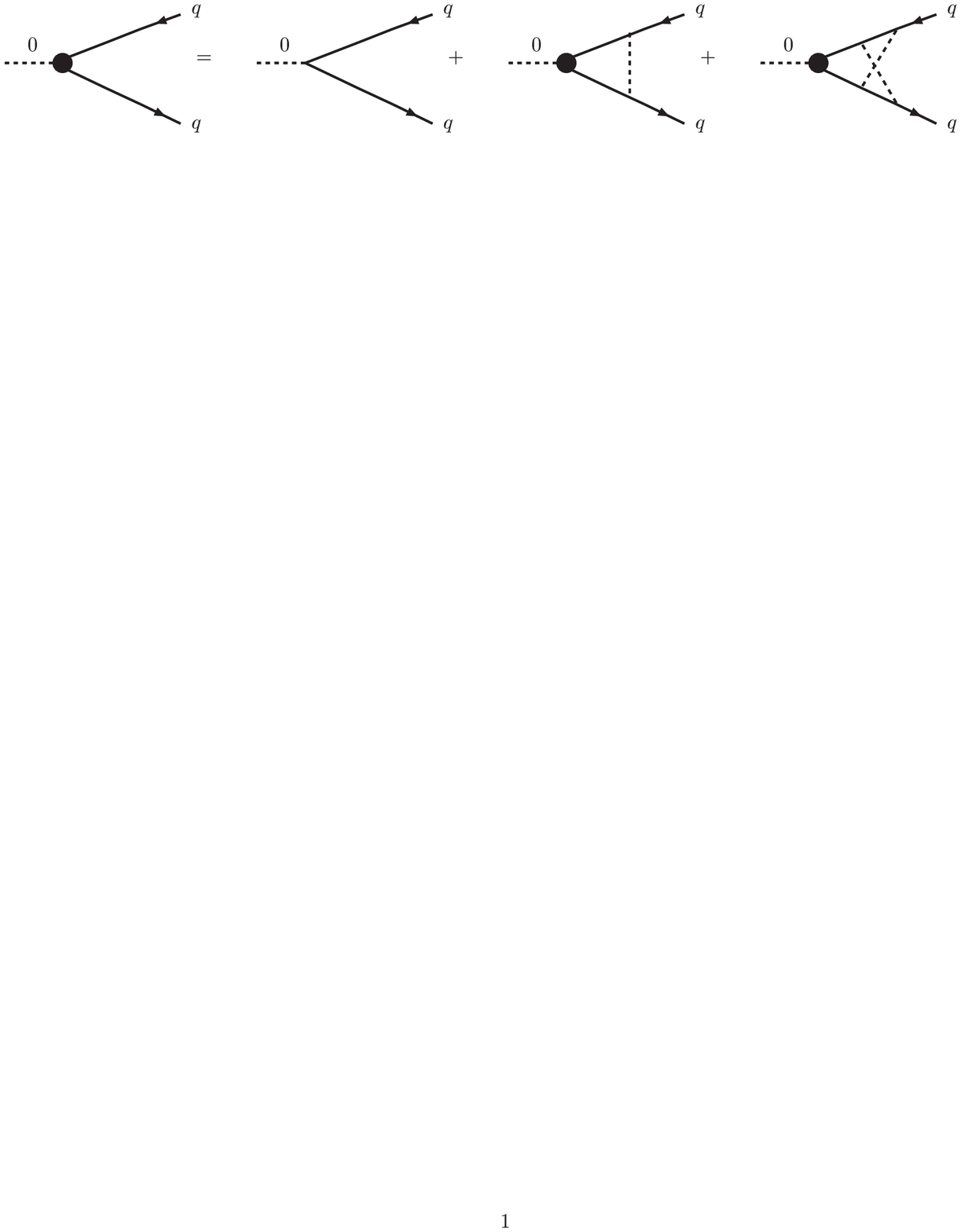}}
\\
\\
where the blob represents the unknown Green function $G_R(a,\ln (-q^2/\mu^2))$.
This linear Dyson--Schwinger equation can be solved by a
scaling solution, $L=\ln (-q^2/\mu^2)$, \be G_R(a,L)=\exp{\{-\gamma_G(a)L\}}.\ee Indeed, this
satisfies the desired normalization and leads to the equation \be
\exp{\{-\gamma_G(a)L\}}=1+
\left(\exp{\{-\gamma_G(a)L\}}-1\right)\left[aF_1(\gamma_G)+a^2F_2(\gamma_G)\right],\ee
where the two Mellin transforms are the functions determined by
\be
F_1: \rho\to
 - \int \frac{d^4k}{i \pi^2} \frac{1}{k^2(k-q)^2} \left( \frac{k^2}{q^2} \right)^{-\rho}
\ee and
similarly
\be
F_2: \rho\to
 \int_{\mathbb{M}} \frac{d^4k}{i \pi^2} \int_{\mathbb{M}} \frac{d^4l}{i \pi^2}
 \frac{l\!\!/(k\!\!/+q\!\!/)}{(k+l)^2 l^2(k+q)^2k^2(l-q)^2} \left[ \frac{(l+k)^2}{q^2} \right]^{-\rho}.
\ee
Clearing the factor
$[\exp\{-\gamma_G(a)L\}-1]$ in this equation gives \be
1=aF_1(\gamma_G)+a^2F_2(\gamma_G).\ee It remains to determine
$F_1,F_2$ explicitely and solve this implicit equation for
$\gamma_G$ in terms of $a$. We do so in the next sections but first
discuss the perturbative structure behind this solution.
\subsection{The algebraic structure}
We can identify any graph in this resummation with a word in  two
letters $u,v$ say, for example:
\be\bkw\ee
 We have renormalized Feynman rules $\phi_R$ such that \be\phi_R(u)(L)=-L\lim_{\rho\to 0}\rho F_1(\rho),\ee
and \be\phi_R(v)(L)=-L\lim_{\rho\to 0}\rho F_2(\rho).\ee

The Green function $G_R(a,L)$ is obtained as the
evaluation by $\phi_R$ of the fixpoint of the combinatorial Dyson Schwinger equation
\be X(a)=\One+aB_+^u(X(a))+a^2B_+^v(X(a)).\ee
We have
\be
X(a)=1+au+a^2(uu+v)+\ldots=\exp_{\curlyvee}{[au+a^2v]},\ee
where $\curlyvee$ is the shuffle \be H_{\rm lin}\times H_{\rm
lin}\to H_{\rm lin},\ee \be B_+^i(w_1)\curlyvee B_+^j(w_2)=B_+^i(w_1\curlyvee B_+^j(w_2))+
B_+^j(w_2\curlyvee B_+^i(w_1)),\ee
$\forall i,j\in\{u,v\}$. Note that for example $u\curlyvee u=2 uu$.

The two maps $B_+^i$ are Hochschild one-cocycles, and $X(a)$ is group-like: \be \Delta X(a)=X(a)\otimes X(a).\ee
Correspondingly, decomposing $X(a)=1+\sum_{k\geq 1}a^k c_k$, we have
\be \Delta c_k=\sum_{j=0}^k c_j\otimes c_{k-j},\ee
with $c_0=1$.
This is a decorated version of the Hopf algebra of undecorated ladder trees $t_n$ with coproduct $\Delta t_n=\sum_{j=0}^n t_j\otimes t_{n-j}$.
 Feynman rules become iterated integrals as
\be \phi_R(B_+^i(w))(L)=\int \left\{\phi(w)(\ln k^2/\mu^2)d\mu_i(k)\right\}_-,\ee
where $d\mu_i$ is the obvious integral kernel for $i\in u,v$, cf.\ (\ref{DSEint}).
Apart from the shuffle product,
we have disjoint union as a product which makes the Feynman rules into a character
\be \phi(w_1\cdot w_2)=\phi(w_1)\phi(w_2).\ee
These two commutative products $\curlyvee,\cdot$ allow to express the primitive elements associated with shuffles of letters $u,v$, see for example \cite{Bigrad}:
\begin{thm}
The primitive elements are given by polarization of the primitive elements $p_n$ of the undecorated ladder trees $t_n$.
These are given by $p_n=\frac{1}{n}[S\star Y](t_n)$.
\end{thm}
Here, $Y$ denotes the grading operator, defined by $Y(t_k) = k t_k$ and the star product is defined as usual
by $O_1 \star O_2 = \cdot \circ ( O_1 \otimes O_2 ) \circ \Delta$.
Polarization of the undecorated primitive elements $p_n$ means that we decorate each vertex of $p_n$ with $u+v$.

The set ${\mathcal P}(u,v)$ of primitive elements is hence spanned by elements $p_{i_u,i_v}$, where the integeres $i_u,i_v$ count the number of letters $u$ and $v$ in the polarization of $t_{i_u+i_v}$.
For example the primitive element corresponding to the undecorated ladder tree $t_2$ is $p_2=t_2-\frac{1}{2}t_1t_1$.
Polarization yields
\bea
 & & p_{2,0} = \frac{1}{2} ( u \curlyvee u - u \cdot u ) = u u - \frac{1}{2} u \cdot u,
 \;\;\;
 p_{0,2} = \frac{1}{2} ( v \curlyvee v - v \cdot v ) = v v - \frac{1}{2} v \cdot v,
 \\
 & & p_{1,1} = u \curlyvee v - u \cdot v = u v + v u - u \cdot v.
 \nonumber
\eea

\section[Mellin]{The Mellin transforms}
The general structure of the Mellin transform can be obtained from
quite general considerations. The crucial input comes from
powercounting and conformal symmetry. \begin{thm} The Mellin transforms above are invariant
under the transformation $\rho\to 1-\rho$.
\end{thm}
Proof: Explicit computation. We give it here for $F_1$. We assume
$\Re\rho>0$ so that $F_i$ is well defined as a function. Then, the
conformal inversion $k_\mu\to k^\prime_\mu=k_\mu/k^2$ gives explicitly
\be
- \int \frac{d^4k^\prime}{i \pi^2}
\frac{1}{{k^\prime}^2(k^\prime-q)^2} \left( \frac{{k^\prime}^2}{q^2} \right)^{-1+\rho}
\ee
for $F_1$. $F_2$ can be treated similarly by conformal inversion in both Minkowski spaces.\hfill $\Box$
\subsection{The Mellin transform of the one-loop kernel}
This Mellin transform is readily integrated to deliver \be F_1(\rho)=
-\frac{1}{\rho(1-\rho)},\ee exhibiting the expected conformal symmetry.
\subsection{The Mellin transform of the two-loop kernel}
Determining this Mellin transform is the core part of this paper. We
proceed by making use of the advantage that we remain in four
dimensions, and use results of \cite{BGK}.
We are interested in the integral
\be
F_2(\rho) =
 (-q^2)^\rho \int_{\mathbb{M}} \frac{d^4k}{i \pi^2} \int_{\mathbb{M}} \frac{d^4l}{i \pi^2}
 \frac{l\!\!/(k\!\!/+q\!\!/) [-(l+k)^2]^{-\rho}}{(k+l)^2 l^2(k+q)^2k^2(l-q)^2}.
\ee
Integration is over the eight dimensional cartesian product of two Minkowski spaces furnished with a quadratic form \be a^2=a_0^2-a_1^2-a_2^2-a_3^2.\ee
A simple tensor calculus delivers
\be
 F_2(\rho)=\frac{1}{2}\left\{-2G_4(1,1+\rho)G_4(1,1+\rho)+I_6(1,1,\rho,1,1,2-\rho)+I_6(1,1,1+\rho,1,1,1-\rho)\right\},
\ee
where \be G_D(a,b)=\frac{\Gamma(a+b-D/2)\Gamma(D/2-a)\Gamma(D/2-b)}{\Gamma(a)\Gamma(b)\Gamma(D-a-b)},\ee
so that $G_4(1,1+\rho)=\frac{1}{\rho(1-\rho)}$. We use the notation of \cite{BGK} for $I_6$.
In this notation, we have $I_6=\overline{I_6}$.
Setting $u\to 0$ and $v=-\rho$ or $v=1-\rho$ we can determine the two $I_6$ integrals as a limit $u\to 0$ in Eq.(19)(op.cit.)
as
\be I_6(1,1,1-v,1,1,1+v)=8\sum_{n=1}^{\infty}n\zeta_{2n+1}(1-2^{-2n})v^{2n-2},\ee
and similarly for $v=1-\rho$.
We hence find the above DSE in the form
\be
1=-a\frac{1}{\gamma_G(1-\gamma_G)}-a^2\left\{ \frac{1}{\gamma_G^2(1-\gamma_G)^2}-4\sum_{n=1}^\infty n\zeta_{2n+1}(1-2^{-2n})\left[\gamma_G^{2n-2}+(1-\gamma_G)^{2n-2}\right]\right\}.\ee
\section{The solution}
We can solve for $\gamma_G$ in the above in two different ways, expressing the solution as an infinite product or via the logarithmic derivative of the $\Gamma$ function.
\subsection{Solution as an infinite product}
We have:
\be G_R(a,L)=\exp{\left\{\sum_{p\in {\mathcal P}(u,v)}a^{|p|}\phi_R(p)(L)\right\}}.\ee
Here, the sum is over all primitives $p\in {\mathcal P}(u,v)$, where ${\mathcal P}$ is the set of primitives assigned to any tree $t_n$ decorated arbitrarily by letters
in the alphabet $u,v$, as described above. The proof is an elementary exercise in the Taylor expansion of the two Mellin transforms.
Note that $\phi_R(p)(L)$ is linear in $L$ for primitive $p$, $\partial^2_L \phi_R(L)=0$. We hence find for $\gamma_G(a)$
\be \gamma_G(a)=-\frac{\partial \ln G}{\partial L}|_{L=0}=-\sum_p a^{|p|}\phi_R(p)/L.\ee
Convergence of the sum is covered by the implicit function theorem, which provides for $\gamma_G$ through the two Mellin transforms in the DSE. We hence proceed to the second way of expressing the solution.
\subsection{Solution via the $\psi$-function}
We can express the DSE using the logarithmic derivative of the $\Gamma$ function and we obtain
\bea
1 & = & -a\frac{1}{\gamma_G(1-\gamma_G)}
  -a^2\left\{
    \frac{1}{\gamma_G^2(1-\gamma_G)^2}
 \right. \\
 & & \left.
    + \frac{1}{\gamma_G} \left[ \psi'\left(1+\gamma_G\right) - \psi'\left(1-\gamma_G\right) \right]
    + \frac{1}{1-\gamma_G} \left[ \psi'\left(2-\gamma_G\right) - \psi'\left(\gamma_G\right) \right]
 \right. \nonumber \\
 & & \left.
    - \frac{1}{2\gamma_G} \left[ \psi'\left(1+\frac{\gamma_G}{2}\right) - \psi'\left(1-\frac{\gamma_G}{2}\right) \right]
    - \frac{1}{2(1-\gamma_G)} \left[ \psi'\left(\frac{3-\gamma_G}{2}\right) - \psi'\left(\frac{1+\gamma_G}{2}\right) \right]
      \right\}.
 \nonumber
\eea
Here
\be
 \psi'(x) = \frac{d^2}{dx^2} \ln \Gamma(x).
\ee
Again, the two-loop solution shows explicitly the conformal symmetry $\gamma_G \rightarrow 1-\gamma_G$.
Note that the apparent second order  poles at $\gamma_G=0$ and $\gamma_G=1$ on the rhs are only first order poles upon using standard properties
of the logarithmic derivative of the $\Gamma$ function, as it has to be.
This provides an implicit equation for $\gamma_G$, which can be solved numerically.
\section{Conclusions}
We determined the Mellin transform of the two-loop massless vertex
in Yukawa theory. We used it to resum a linear Dyson--Schwinger
equation. Following \cite{BKDSE,dktor,KY}, more complete
applications to non-linear Dyson--Schwinger equations will be
provided elsewhere. Techniques to deal with non-linearity have
indeed been developed recently \cite{BKDSE,dktor,KY}, and involve
transcendental functions even upon resummation of terms from the
first Mellin transform \cite{BKDSE}. In the non-linear case one gets
indeed results very different from scaling, as has been demonstrated
early on in field theory \cite{Roman}. Finally, we note that the
same two-loop Mellin transform also appears in setting up the full
DSE in other renormalizable theories \cite{KY2}.
\section*{Acknowledgments} We thank Roman Jackiw for pointing his thesis \cite{Roman} out
to us.

\end{document}